\documentclass[twocolumn,aps,prl,showpacs]{revtex4}

\usepackage{amsmath} 
\usepackage{graphicx}
\usepackage{bm}
\usepackage{amsfonts}

\begin{document}

\title{Emergence of Long-range Correlations and Rigidity at the Dynamic Glass Transition}

\author{Grzegorz Szamel and Elijah Flenner}

\affiliation{Department of Chemistry, 
Colorado State University, Fort Collins, CO 80523
}

\date{\today}

\pacs{61.43.Fs, 05.20.-y, 64.70.Q-}

\begin{abstract}
At the microscopic level, 
equilibrium liquid's translational symmetry is spontaneously broken at
the so-called dynamic glass transition predicted by
the mean-field replica approach. We show that this fact 
implies the emergence of Goldstone modes and
long-range density correlations. We derive and evaluate a new 
statistical mechanical expression for the glass shear modulus.  
\end{abstract}
\maketitle

\textit{Introduction.} --- The difference between a liquid and a solid manifests 
itself in their response to a volume preserving shear deformation: a liquid
flows whereas in a solid internal forces arise which are proportional to and 
oppose the deformation. Solid's rigidity is an emergent \cite{PWAS} property: 
although it amounts to transmitting internal forces opposing 
a volume preserving deformation over macroscopic distances, it does not
originate from long-range intermolecular interactions. However, it is intuitively
plausible, and it was argued at a phenomenological level \cite{Forster,Raveche},  
that the emergence of rigidity has to be accompanied by appearance of 
long-range correlations. 

Interestingly, the first \emph{microscopic} argument for the existence of long-range 
correlations in \emph{crystalline} solids was given in 1966 by Wagner \cite{Wagner}, 
preceding phenomenological analyzes of Refs. \cite{Forster,Raveche}. 
However, the part of his paper dealing with solids seems to have been overlooked. 
More importantly, while he showed that in crystalline solids broken translational 
symmetry implies long-range correlations, he did not connect the latter to 
solids' rigidity.

The intimate relation between long-range correlations and rigidity was
established somewhat indirectly. Bavaud \textit{et al.} \cite{Bavaud} 
analyzed the standard expression for the elastic moduli, which is derived
by calculating free energy change due to a deformation. They proved that this 
expression reduces itself to 
the inverse of the isothermal compressibility if both intermolecular interactions 
and density correlation functions are short range. Thus, they showed
that for systems with short-range interactions the absence of long-range density
correlations implies the vanishing of the shear modulus. 

The microscopic argument for the existence of long-range correlations in crystals
%crystalline solids 
was independently derived by Szamel and Ernst \cite{SE}. They proposed an
expression for a displacement field in crystals in terms of the microscopic density
and %. They 
showed that the displacement field defined in this way exhibits long-range 
correlations, as expected on the basis of phenomenological arguments 
\cite{Forster,Raveche}. 

Here we consider an \textit{amorphous} solid which appears as a consequence of 
the so-called dynamic glass transition \cite{KTW} predicted by the mean-field
replica approach \cite{FP,CFP,MP}. We show that, in close analogy with crystalline
solids, broken translational symmetry of this amorphous solid implies 
the emergence of long-range correlations and rigidity. 

It should be recalled that, although macroscopically the amorphous solid's 
density is constant, microscopically it is randomly non-uniform. Thus, 
a rigid translation of the amorphous solid produces an equivalent but 
different state. To describe this randomly broken translational symmetry, 
and amorphous solids in general, we use the replica approach of Franz and Parisi 
\cite{FP,CFP}. In this approach, the appearance of the amorphous solid 
manifests itself in non-trivial replica off-diagonal densities.
We argue that the existence of distinct, rigidly shifted states of the amorphous solid  
manifests itself in a family of equivalent sets of replica off-diagonal densities.
Next, we propose an expression for the displacement field in terms
of microscopic two-point replica off-diagonal densities. 
We show that the displacement field defined in this way 
exhibits long-range correlations. Since the displacement field is defined 
in terms of microscopic two-point replica off-diagonal densities, 
long range of the displacement field's 
correlations implies a slow decay of a four-point replica 
off-diagonal correlation function. Finally, 
we derive and evaluate a new expression for the 
shear modulus of amorphous solids. To this end we evaluate at the microscopic
level the force needed to maintain a shear deformation and compare it to 
an expression obtained from the macroscopic theory of elasticity.

\textit{Family of equivalent glassy states.} ---
Following Franz and Parisi \cite{FP,CFP} we consider a system
of particles coupled to a frozen (referred to as
``quenched'') configuration of the same system via an attractive potential 
\begin{equation}\label{attrorig} 
\epsilon \sum_{i,j} w(|\mathbf{r}_i-\mathbf{r}_j^0|),
\end{equation}
where $\mathbf{r}_i$ and $\mathbf{r}_j^0$ are positions of particles
$i$ and $j$ in the system and in the quenched configuration, respectively,
and $w(r)$ is a monotonic function with a minimum at $r=0$. We assume that
when the amplitude $\epsilon$ of the potential is gradually reduced, if the 
temperature is low enough or if the density is high enough, 
the system gets stuck in a metastable state characterized by non-trivial 
correlations with the quenched configuration and that this
metastable state survives in the $\epsilon\to 0$ limit. 
The appearance of such a metastable state is termed
the dynamic glass transition \cite{FP,CFP,MP}. To facilitate 
averaging over a distribution of quenched configurations we replicate the system
$s$ times.  Including the quenched configuration as the $0$th replica we %then 
have $m=s+1$ times replicated the system and at the end we
take the limit $s\to 0$ or $m\to 1$ \cite{comment1}.
At the level of the replicated system, 
the dynamic glass transition manifests itself in the appearance
of non-trivial two-point replica off-diagonal densities $n_{\alpha\beta}(r)$,
$\alpha\neq\beta$. With $w(r)$  
having a minimum at $r=0$, these functions exhibit a pronounced peak
at $r=0$, oscillate in phase with the equilibrium pair distribution 
function $g(r)$ and decay to $n^2$ at large distances ($n$ is the average density).
In the following the metastable state with all replica off-diagonal densities
having the main peak at $r=0$ will be referred to as the ``classical'' state.
 
The above
described construction can also be performed with a potential that pins the system 
in a position shifted with respect to the quenched configuration,
\begin{equation}\label{attrshift} 
\epsilon \sum_{i,j} w(|\mathbf{r}_i-\mathbf{r}_j^0-\mathbf{a}|).
\end{equation}
The metastable states obtained with potentials \eqref{attrorig} and 
\eqref{attrshift} are identical up to a rigid shift by vector $\mathbf{a}$ of the system
relative to the quenched configuration.
At the level of the replicated system 
this translation corresponds to the following 
transformation of two replica densities,
\begin{equation}\label{nabshift}
n_{\alpha 0}(r) \rightarrow n_{\alpha 0}(|\mathbf{r}-\mathbf{a}|), \;\;
n_{0\alpha}(r) \rightarrow n_{0\alpha}(|\mathbf{r}+\mathbf{a}|)
\end{equation}
for $\alpha >0$ and 
keeping all other two replica densities (both diagonal and off-diagonal) unchanged. 
In the following we consider only densities $n_{\alpha 0}$, $\alpha >0$, 
since densities $n_{0 \alpha}$ can be obtained from the former ones.

This observation implies that instead of one glassy state we have a continuous 
family of states that can be labeled by a vector $\mathbf{a}$ denoting translation 
with respect to the ``classical'' state.
%, \textit{i.e.} the state with 
%$n_{\alpha 0}(r)$, $\alpha>0$, having the main peak at $r=0$. 
This is 
analogous to what is found in other systems with broken continuous symmetry,
\textit{e.g.} in ferromagnets \cite{Forster} or in crystals \cite{SE}. 

We note that the rigid shift described above breaks replica symmetry. However, 
since replica non-symmetric states obtained by rigid translations arise due to
broken translational symmetry, they have the same free energy as the 
``classical'' solution (which is replica symmetric \cite{FP,CFP}). Thus, the present
case is fundamentally different from replica symmetry breaking in mean-field spin 
glasses \cite{ParisiRSB}. In the latter case, replica symmetry breaking originates
from the existence of a multitude of different glassy states unrelated by 
translations. Importantly, the replica non-symmetric state minimizes the free energy.

\textit{Microscopic definition of displacement field.} --- According to the conventional 
definition, a displacement field in crystalline solids is 
defined in terms of departures of particles' instantaneous positions from 
their lattice sites. Notably, 
this definition implicitly assumes that each particle can be assigned to 
a specific lattice site. Thus, it is not applicable for, \textit{e.g.} crystals
with vacancies or interstitials. This fact was one of the motivations for 
a microscopic  definition of the displacement field \cite{SE,WF}.
%a definition of the displacement field in terms of the microscopic density which was
%proposed in Refs. \cite{SE,WF}.

Since amorphous solids do not have any underlying crystalline lattice, it is 
impossible to use the conventional definition. 
An operational definition that is closest to the one 
used for crystals is as follows. First, average positions of particles
are identified. Then, the displacement field is defined in terms of 
departures of the 
instantaneous positions with respect to the average positions. While
a procedure like this can be
%, at least in principle, 
implemented in a computer
simulation \cite{MKK} or an experiment \cite{Ghosh}, it is not clear how to
formulate it theoretically. The definition we propose is similar in 
spirit to this procedure 
in that it relates microscopic and average densities of the amorphous solid.

We propose the following definition of (the Fourier transform of) the 
displacement field in terms of microscopic replica off-diagonal density, 
\begin{eqnarray}\label{disdef}
\mathbf{u}(\mathbf{k}) &=& - \frac{1}{{\cal{N}}s} \int d\mathbf{r}_1 
e^{-i\mathbf{k}\cdot\mathbf{r}_1} \int d\mathbf{r}_{21}
\sum_{\alpha>0}
\frac{\partial n_{\alpha 0}(\mathbf{r}_1,\mathbf{r}_2)}{\partial \mathbf{r}_1} 
\nonumber \\ && \times
\sum_{i,j}\delta(\mathbf{r}_1-\mathbf{r}_i^{\alpha})\delta(\mathbf{r}_2-\mathbf{r}_j^0).
\end{eqnarray}
Here 
$ n_{\alpha 0}(\mathbf{r}_1,\mathbf{r}_2)$ is the ``classical'' two-point replica 
off-diagonal density and
$\sum_{i,j}\delta(\mathbf{r}_1-\mathbf{r}_i^{\alpha})\delta(\mathbf{r}_2-\mathbf{r}_j^0)$
is the \textit{microscopic} two-point replica off-diagonal density 
with $\mathbf{r}_i^{\alpha}$ and $\mathbf{r}_j^0$ being positions of particles $i$ and
$j$ in replica $\alpha$ and $0$, respectively.
Finally, $\cal N$ is the normalization factor, 
${\cal{N}} = (3s)^{-1} \int d\mathbf{r}_{21} \sum_{\alpha>0}
\left(\partial_{\mathbf{r}_1} n_{\alpha 0}(\mathbf{r}_1,\mathbf{r}_2)\right)^2$.
Note that definition \eqref{disdef} is symmetric with respect
to replica indices $\alpha$, $\alpha>0$. This reflects the fact
that the system is first deformed and only subsequently replicated. Thus, 
deformations in all $\alpha >0$ replicas are the same.

Importantly, definition \eqref{disdef} is only applicable 
to infinitesimally small deformations of the
``classical'' state. A similar restriction applies to the definition of
the displacement field in crystals proposed in Refs. \cite{SE,WF}.

\textit{Long-range density correlations.} --- 
First, we adapt the calculation presented in Ref. \cite{SE} and
prove that correlations of the displacement field defined through Eq. \eqref{disdef}
exhibits a small wavevector divergence.  
Then we show that long-range displacement
field correlations imply slow decay of a component of a four-point replica off-diagonal
correlation function.

We start with Bogoliubov's inequality \cite{Forster,comment2}
$\left< |A|^2 \right>\left< |B|^2 \right> 
\geq |\left< A B \right>|^2 
$
with 
$A = V^{-1/2} \hat{\mathbf{n}}\cdot\mathbf{u}^*(\mathbf{k})$
and 
$B= V^{-1/2} s^{-1} 
\sum_{\alpha>0} \hat{\mathbf{n}}\cdot i {\cal{L}}_{\alpha}\mathbf{g}_{\alpha}(\mathbf{k})
$
where $\hat{\mathbf{n}}$ is an arbitrary unit vector, 
$\mathbf{g}_{\alpha}$ is the Fourier transform of the 
microscopic momentum density in replica $\alpha$,
$\mathbf{g}_{\alpha}(\mathbf{k}) = \sum_i m \mathbf{v}_i^{\alpha} 
e^{-i\mathbf{k}\cdot\mathbf{r}_i^{\alpha}},
$
with $m$ denoting the particle's mass and 
$\mathbf{v}_i^{\alpha}$ being the velocity of particle $i$ in replica, 
and ${\cal{L}}_{\alpha}$ is the Liouville operator \cite{Forster} in replica $\alpha$.

The cross term can be evaluated using the  self-adjoint property of 
${\cal{L}}_{\alpha}$. In the $\mathbf{k}\to 0$ limit we obtain
$\left< A B \right> = - k_B T/s$.
$\left< |B|^2 \right>$ can be expressed in terms of the correlation 
function of the stress tensor. The stress tensor in replica $\alpha$ is 
defined through the continuity equation for the momentum in replica $\alpha$,
$i{\cal{L}}_{\alpha} \mathbf{g}_{\alpha}(\mathbf{k}) = - i \mathbf{k}\cdot
\stackrel{\leftrightarrow}{\sigma}_{\alpha}(\mathbf{k};t)$.

Substituting $\left< AB\right>$ and $\left< |B|^2 \right>$ into Bogoliubov's inequality,
and taking the $s\to 0$ limit we obtain
\begin{widetext}

\vspace*{-.2in}
\begin{equation}\label{udiv}
\lim_{s\to 0} 
\frac{s}{V} \left< |\hat{\mathbf{n}}\cdot\mathbf{u}(\mathbf{k})|^2 \right>
\geq \frac{1}{k^2}
\frac{\left(k_B T\right)^2}
{\lim_{\mathbf{k}\rightarrow 0}\frac{1}{V} 
\left< |\hat{\mathbf{k}}\cdot\stackrel{\leftrightarrow}{\sigma}_1(\mathbf{k})
\cdot\hat{\mathbf{n}}|^2 - 
|\hat{\mathbf{k}}\cdot\stackrel{\leftrightarrow}{\sigma}_1(\mathbf{k})
\cdot\hat{\mathbf{n}}|
|\hat{\mathbf{k}}\cdot\stackrel{\leftrightarrow}{\sigma}_2(\mathbf{k})
\cdot\hat{\mathbf{n}}|
\right>}
\end{equation}
\end{widetext}
where  $\hat{\textbf{k}}=\textbf{k}/k$. 

Bound \eqref{udiv} for the small wavevector
behavior of the displacement field correlations is compatible with the result
of a phenomenological analysis which predicts a $k^{-2}$ divergence \cite{Forster}. 
This agreement, and the closeness of the steps outlined above and
the general scheme presented in Chapter 7 of Ref. \cite{Forster} 
provide an additional justification
for definition \eqref{disdef} and for recognizing the displacement field as a
``hydrodynamic Goldstone mode''.

Finally, since the displacement field is defined in terms of microscopic two-point
replica off-diagonal densities, 
the left-hand-side (LHS) of Eq. \eqref{udiv} can be re-written
in terms of four-point replica off-diagonal densities,
\begin{eqnarray}\label{n4div}
&& \!\!\!\!\! \text{LHS}
= \frac{1}{{\cal{N}}V}
\int d\mathbf{r}_1 ... d\mathbf{r}_4
\hat{\mathbf{n}}\cdot
\frac{\partial n_{10}(\mathbf{r}_1,\mathbf{r}_2)}{\partial\mathbf{r}_1}
\hat{\mathbf{n}}\cdot
\frac{\partial n_{10}(\mathbf{r}_3,\mathbf{r}_4)}{\partial\mathbf{r}_3}
\nonumber \\ && \!\!\!\!\! \times
\left( n_{1010}(\mathbf{r}_1,\mathbf{r}_2,\mathbf{r}_3,\mathbf{r}_4)
- n_{1020}(\mathbf{r}_1,\mathbf{r}_2,\mathbf{r}_3,\mathbf{r}_4) 
\right) e^{-i\mathbf{k}\cdot\mathbf{r}_{13}}.
\end{eqnarray}

The combination of Eqs. \eqref{udiv} and \eqref{n4div} proves the 
slow decay of a component of a four-point replica off-diagonal correlation function
of the amorphous solid. 
An analogous calculation for a crystal proves the slow decay of 
correlations of high Fourier components of the (non-translationally invariant)
two-particle density \cite{Wagner,SE,WF}.

\textit{Shear modulus.} --- The derivation of a new formula for the amorphous solid's 
shear modulus proceeds in two steps. First, we consider the change of replica 
off-diagonal densities upon an infinitesimally small, long wavelength
deformation. Next, we calculate microscopically the force needed to maintain such a 
deformation. Comparison of this force with the corresponding macroscopic 
expression allows us to identify the shear modulus.

According to Eq. \eqref{nabshift}, under an infinitesimally small
uniform translation, replica off-diagonal densities 
$n_{\alpha 0}$, $\alpha >0$, change in the following way:
\begin{equation}\label{habshiftinfin}
n_{\alpha 0}(r) \rightarrow 
n_{\alpha 0}(r) -\mathbf{a}\cdot\partial_{\mathbf{r}}n_{\alpha 0}(r).
\end{equation}

To generalize Eq. \eqref{habshiftinfin} to an infinitesimally small non-uniform
translation we imagine a pinning potential which deforms the system by 
imposing an infinitesimal shift that depends on the position of the particle.
This suggests the following generalization of Eq. \eqref{habshiftinfin},
\begin{eqnarray}\label{habdefinfin}
\lefteqn{ 
n_{\alpha 0}(r_{12}) \to 
n_{\alpha 0}(\mathbf{r}_1,\mathbf{r}_2) }  
\nonumber \\ && =   
n_{\alpha 0}(r_{12}) 
-\mathbf{a}(\mathbf{r}_1)\cdot\partial_{\mathbf{r_1}}
n_{\alpha 0}(r_{12}),
\end{eqnarray}
where $r_{12}=|\mathbf{r}_1-\mathbf{r}_2|$ and 
$\mathbf{a}(\mathbf{r}_1)$ is an infinitesimal, long-wavelength 
deformation imposed on the system.
We emphasize that \eqref{habdefinfin} is only assumed for
a long wavelength volume-preserving deformation. Specifically, we assume a transverse 
deformation such that 
$\partial_{\mathbf{r}}\cdot\mathbf{a}(\mathbf{r})=0$. 

Eq. \eqref{habdefinfin} is inspired by the assumption made
by Triezenberg and Zwanzig \cite{TZ}, and by Lovett \textit{et al.} \cite{LDeHVB} 
in their analyzes of surface tension. 
A similar assumption was made by Szamel and Ernst \cite{SE}
in their derivation of shear modulus of crystalline solids. Finally, 
assumption \eqref{habdefinfin} is consistent with our microscopic definition
of the displacement field: if all densities $n_{\alpha 0}$, $\alpha>0$, are 
changed as indicated in \eqref{habdefinfin} then the Fourier transform of the
average displacement field is equal to 
$\left<\mathbf{u}(\mathbf{k})\right> = 
\int d\mathbf{r}
e^{-i\mathbf{k}\cdot\mathbf{r}} \mathbf{a}(\mathbf{r}).
$

Next, we calculate the force exerted on one replica of the system by the pinning 
potential chosen in such a way that it imposes the replica 
off-diagonal densities \eqref{habdefinfin}. 

First, we identify the pinning potential needed to maintain 
densities \eqref{habdefinfin}. 
Since we are only interested in infinitesimally small
deformations and thus infinitesimally small changes of replica off-diagonal densities,
we resort to a (static) linear-response type relation. Thus, we express 
the pinning potential using a functional
derivative of the inter-replica potential with respect to the replica off-diagonal
density,
\begin{equation}\label{pinning}
\sum_{\beta>0} \int d\mathbf{r}_3 d\mathbf{r}_4
\left(\frac{ \delta V_{\alpha 0}(\mathbf{r}_1,\mathbf{r}_2) }
{ \delta n_{\beta 0}(\mathbf{r}_3,\mathbf{r}_4) }\right)_n
\left[- \mathbf{a}(\mathbf{r}_3)\cdot \partial_{\mathbf{r_3}}
n_{\beta 0}(r_{34})\right].
\end{equation}
Note that since we have assumed a volume-preserving deformation, 
the functional derivative in expression \eqref{pinning} should be taken 
at constant density.

Then, we calculate the force per unit volume exerted by the above pinning
potential on the replica $\alpha$. The Fourier transform of this force,
after integrating by parts and using the transverse character of the deformation,
can be written in the following way:
\begin{eqnarray}\label{forcek}
\mathbf{F}_{\alpha}(\mathbf{k}) &=& 
- \frac{1}{V} \int d\mathbf{r}_1 ... d\mathbf{r}_4
e^{-i\mathbf{k}\cdot\mathbf{r}_{13}}
\left(\partial_{\mathbf{r}_1} n_{\alpha 0}(r_{12})\right)
\\ \nonumber && \times
\sum_{\beta}
\left(\frac{\delta V_{\alpha 0}(\mathbf{r}_1,\mathbf{r}_2)}
{\delta n_{\beta 0}(\mathbf{r}_3,\mathbf{r}_4)} \right)_{n}
\left(\partial_{\mathbf{r}_3} n_{\beta 0}(r_{34})\right)
\cdot \mathbf{a}(\mathbf{k}).
\end{eqnarray}

Finally, we take the long wavelength limit and expand the right-hand-side of
Eq. \eqref{forcek} in a power series in $\mathbf{k}$. The zeroth order term
vanishes because no force is needed in order to accomplish a rigid shift 
of the amorphous solid. 
The first order term vanishes by symmetry. The second order term, in the $s\to 0$ limit
and for a shear deformation involving $y$-dependent translation along the $x$ axis, 
can be written in the following form
\begin{equation}\label{second}
F_x(k_y) = \mu k_y k_y a_x(k_y)
\end{equation}
where $\mu$ is given by
\begin{eqnarray}\label{mu}
&& \!\!\!\!\!  
\mu = \frac{k_B T}{2V} \int d\mathbf{r}_1 ... d\mathbf{r}_4 \;
y_{13}^2
\left(\partial_{x_1}n_{10}(r_{12})\right)
\left(\partial_{x_3}n_{10}(r_{34})\right)
\nonumber \\ 
&& \!\!\!\!\!  
\times
\left(\left(\frac{\delta(\beta V_{10}(\mathbf{r}_1,\mathbf{r}_2))}
{\delta n_{10}(\mathbf{r}_3,\mathbf{r}_4)} \right)_{n}
- \left(\frac{\delta(\beta V_{10}(\mathbf{r}_1,\mathbf{r}_2))}
{\delta n_{20}(\mathbf{r}_3,\mathbf{r}_4)} \right)_{n} \right).
\end{eqnarray}

Macroscopically, for an isotropic solid 
the force needed to maintain a long wavelength shear deformation
is given by a formula identical to Eq. \eqref{second}, where $\mu$ is the
shear modulus. This fact allows us to identify $\mu$ given by Eq. \eqref{mu}
as the shear modulus of the amorphous solid. 

To evaluate $\mu$ one has to use a concrete (albeit necessarily approximate) 
implementation of the replica approach in order 
to calculate replica off-diagonal densities and functional derivatives in 
Eq. \eqref{mu}. To obtain $n_{\alpha 0}(r)$ we use the recently
proposed version \cite{GSreplica} which is consistent
with mode-coupling theory \cite{Goetze}.
To calculate the derivatives we use an additional approximation, 
\begin{equation}\label{extraapp}
n^2 \delta c_{\alpha 0}(\mathbf{r}_1,\mathbf{r}_2) =
- n_{\alpha 0}(\mathbf{r}_1,\mathbf{r}_2) \beta V_{\alpha 0}(\mathbf{r}_1,\mathbf{r}_2).
\end{equation}
Here $\delta c_{\alpha 0}$ is the change of the replica off-diagonal direct
correlation function due to a weak replica off-diagonal potential 
$V_{\alpha 0}$. Approximation \eqref{extraapp} combined with replica 
Ornstein-Zernicke equations allow us to get explicit
expressions for functional derivatives that enter Eq. \eqref{mu}.

\begin{figure}
\includegraphics[width=2.7in]{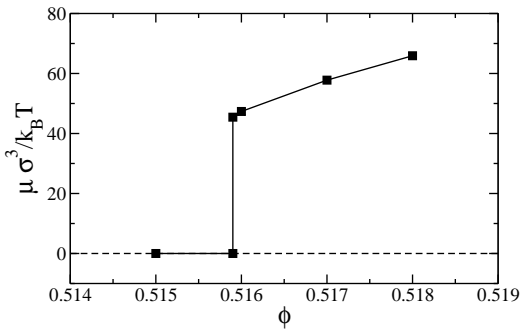}
\caption{\label{fig1}
Dimensionless shear modulus, $\mu\sigma^3/k_BT$, for the hard sphere
glass as a function of the volume fraction, $\phi=n\pi\sigma^3/6$. 
The shear modulus changes discontinuously at the dynamic transition 
which occurs at $\phi_c=0.5159$.}
\end{figure}

In Fig. \ref{fig1} we show the shear modulus for the hard sphere glass 
calculated using as an equilibrium input the Percus-Yevick structure factor.

\textit{Discussion.} --- We identified Goldstone modes and long-range correlations
appearing in the amorphous solid due to the spontaneously broken translational
symmetry. We derived a new expression for the shear modulus of the amorphous
solid. This expression is complementary to the standard one \cite{Bavaud}. 
In particular, our expression can be used to evaluate the shear modulus of the 
hard sphere glass whereas the standard formula is not applicable for hard sphere systems.
In contrast to the result obtained in a recent replica approach study \cite{YM}  
(which was based on the standard formula)
we found a discontinuous change of the shear modulus at the dynamic glass transition. 

We gratefully acknowledge the support of NSF Grant CHE 0909676. GS thanks  
Yukawa Institute, Kyoto University, 
where this work was completed, for its hospitality.


\begin{thebibliography}{99}
\bibitem{comment1} Averaging over an equilibrium distribution of quenched
configurations is implicitly assumed. This results in all replicas being formally
equivalent. Metastable state can be chosen to be replica symmetric \cite{FP,CFP}.
\bibitem{comment2} The brackets $\left< ... \right>$ denote averaging in the
metastable state. We implicitly assume that we can use the usual statistical 
mechanical considerations in this state.
\end{thebibliography}
\end{document}